\documentclass[manuscript]{acmart}

\usepackage{booktabs}
\usepackage{float}
\usepackage{arydshln}
\AtBeginDocument{%
  }

\setcopyright{acmlicensed}
\copyrightyear{2027}
\acmYear{2027}
\acmConference[CHI '27]{CHI Conference on Human Factors in Computing Systems}{2027}{}
\acmISBN{}
\acmDOI{}

\begin{document}

\title{Faithful Where It Can Be Checked: Auditing a Reflection Agent Against Its System Prompt in a Randomized Trial}

\author{Subigya K. Nepal}
\orcid{0000-0002-4314-9505}
\email{sknepal@virginia.edu}
\affiliation{%
  \institution{University of Virginia}
  \city{Charlottesville}
  \state{Virginia}
  \country{USA}
}

\author{Serena Soh}
\affiliation{%
  \institution{Stanford University}
  \city{Stanford}
    \state{California}
  \country{USA}
}
\author{Noah Vinoya}
\affiliation{%
  \institution{Stanford University}
  \city{Stanford}
    \state{California}
  \country{USA}
}
\author{SoHyun Park}
\affiliation{%
  \institution{NAVER Cloud}
  \city{Seongnam}
  \country{Republic of Korea}
}

\author{Mahnaz Roshanaei}
\affiliation{%
  \institution{Stanford University}
  \city{Stanford}
    \state{California}
  \country{USA}
}
\author{Gabriella Harari}
\affiliation{%
  \institution{Stanford University}
  \city{Stanford}
    \state{California}
  \country{USA}
}

\renewcommand{\shortauthors}{Nepal et al.}

\begin{abstract}

Conversational agents are increasingly used to guide reflection. A recent randomized trial compared a GPT-4o career reflection agent with the same program in a static journaling survey. Agent participants ended less committed to their career plans and more doubtful. We coded all 17,930 turns from its two studies, checked our coding against  human coders and linked conversations to the trial's surveys. The rules the agent followed were the easy-to-check ones, like a reply length cap. Told not to flatter, it praised participants in half of its turns; told to challenge gently, it almost never did, and such a break leaves no visible trace. The behavior tied to the worse outcome was the demand to decide: the survey posed each decision once, while the agent asked again when participants hesitated, and those pressed most ended most doubtful. Our findings inform reflection agent design and the writing of checkable instructions.

\end{abstract}



\maketitle

\section{Introduction}

Technology for guided reflection has been a Human-Computer Interaction (HCI) concern for more than a decade, and the arrival of large language models has changed what such technology can do. Earlier systems asked scripted questions about mood logs and activity data like step counts \cite{kocielnik2018}, guided by a design literature on what makes reflection productive \cite{baumer2015, fleck2010, slovak2017}. Newer systems hold open conversations: they support psychiatric patients' daily journaling \cite{kim2024mindfuldiary}, co-write journal sentences alongside their users \cite{kim2024diarymate}, generate adaptive questions for whatever personal challenge a user brings \cite{song2025exploreself} and tailor reflective prompts to behavior sensed from a phone \cite{nepal2024mindscape}. The appeal of conversation over a static reflection form is easy to see. A form accepts whatever the user types, whereas an agent can notice a thin answer and ask for more, remember what was said the day before and pull the threads together at the end. So far, studies of these systems mostly support this appeal: users report feeling engaged and say the conversations deepen their reflection \cite{kim2024mindfuldiary, song2025exploreself}.

Underneath every one of these reflection agents sits a system prompt: a written set of instructions telling the model what to do and how to behave. Some of those instructions set targets anyone can check, like a cap on reply length. Others describe a manner, like being warm but not overly agreeable, and these are the instructions that carry a reflection tool's therapeutic intent. Designers currently have no way to know whether the second kind of instruction is being followed. Nothing in a running system announces that a rule about manner is being broken, and the benchmarks that test instruction following restrict themselves to rules that can be scored automatically \cite{zhou2023ifeval}, which leaves the rules about manner unmeasured. Health intervention research would call this an unverified treatment. Checking that people received the treatment the design intended is basic practice in that field \cite{bellg2004}, and for deployed agents the check has been missing.

Auditing a deployment's transcripts against its own prompt would close this gap. It would show which instructions the agent followed and which it did not, and, if outcomes were also measured, whether the agent behaviors the prompt produced actually helped the people using the system. This requires data that are hard to get: complete conversation logs, a randomized comparison so that outcomes can be interpreted and ideally the same model running more than one prompt, so that a failure can be traced to the model or to the wording. A recent randomized trial of AI-guided career reflection collected all three, and this paper analyzes its complete conversation record.

In the trial, Soh et al.~\cite{soh2026tmb} had emerging adults complete a four-day career reflection program either in conversation with a GPT-4o agent or by answering nearly identical reflection questions in a static survey interface, first with  community college and university students (N = 165) and then with a national sample recruited through Prolific and matched to United States census benchmarks (N = 277 recruited participants, 224 completed the full trial). Daily surveys showed the program working in both formats. At post-test, however, the picture was more mixed. In the larger study, agent-condition participants scored lower than journalers on exploration in depth ($\beta=-.08$), identification with commitment ($\beta=-.07$), career commitment-making ($\beta=-.07$) and commitment making ($\beta=-.05$), and higher on career doubt ($\beta=.04$). 

Two differences ran the other way: agent participants scored higher on career flexibility ($\beta=.06$) and, at the one-month follow-up, on career exploration in breadth ($\beta=.13$), differences the trial reports without interpreting whether they are good or bad for users and reflection systems. The trial also found broader wellbeing differences pointing the same direction as the identity results: the daily gains in positive affect that the program produced were weaker in the agent condition, and agent participants ended the study with lower eudaimonic wellbeing, the sense that one's life has purpose and meaning ($\beta=-.08$). The overall pattern is therefore not that conversation failed but that it tilted participants toward openness and away from settling on their career goals, within a a development program designed to help them explore and commit to a career path. Although these effects are modest in standardized terms, we believe they need explaining because reflection tools are reaching very large user populations, where even small per-person effects add up. Field studies of LLM reflection tools typically report users feeling more engaged and more deeply reflective \cite{kim2024mindfuldiary, song2025exploreself}, so the literature would have predicted the opposite result. There is also a practical reason: the journaling condition costs almost nothing to run, so if talking with an agent is just as, or even less effective than filling out a static reflection form, there is little reason to build the agent.

Why would talking with an agent leave people less settled than reflecting independently in a static journaling form? The trial itself proposes an explanation, grounded in self-determination theory: writing alone may give people a stronger sense of autonomy and ownership over their reflections than answering to an agent does \cite{soh2026tmb, ryan2000}, a reading supported by a small pilot in which a chatbot built on self-determination theory reduced career decision difficulties more than a standard one\cite{han2025}. 

What the trial did not do is look inside the conversations to see whether, and how, the agent actually took autonomy away, because it measured outcomes rather than process. That evidence has to come from the conversations themselves, and other explanations are plausible too. One possibility is that the agent drifted from its instructions in ways nobody observed, which remains an open issue across this literature because trials of conversational agents have rarely analyzed their transcripts \cite{heinz2025, laranjo2018}. Another is that the agent flattered participants so much that it smoothed away the discomfort that honest deliberation requires, a risk that laboratory research on sycophancy has documented \cite{cheng2026, cheng2025elephant}. Conversation may instead have kept participants exploring past the point at which a journaler would have settled, which is what identity theory would predict \cite{luyckx2008} . Or the conversations unfolded as expected, and the difference came from the conversational format itself - the comparison of dyadic interaction with an agent vs. independent self-reflection.
Each account implies a different design response, and choosing among them requires reading the conversations. To our knowledge, no study has systematically coded the conversations of a deployed reflection agent and connected them to the outcomes of a randomized trial.

This paper takes that step. We analyze the trial's full conversation corpus of 17,930 turns, contributed by 185 agent-condition participants across 687 sessions in the two studies. We coded every agent turn for nine conversational moves, among them questions, validation, challenge and demands for a decision, adapting coding schemes from counseling process research \cite{hill2014, chiu2024}. We coded every participant turn for how the person talked about their plans, how certain they sounded and whether they expressed insight, drawing on client change-talk coding from motivational interviewing \cite{miller2013, amrhein2003}. Two human coders first showed that the scheme can be applied reliably. We then validated language model annotators against the human labels, field by field, following emerging practice for LLM annotation \cite{ziems2024, gilardi2023, tornberg2023}, and used the validated annotators to label the full corpus. Because the trial also collected evening surveys, a post-test and a one-month follow-up, we could ask whether anything in the conversations was connected to how participants felt. We ask four questions:

\begin{itemize}
\item \textbf{RQ1.} \textit{How faithfully did the deployed agent follow its system prompt, and where did it deviate?}
\item \textbf{RQ2.} \textit{What did the agent do, turn by turn, and how did that vary across activities and participants?}
\item \textbf{RQ3.} \textit{How did participants' commitment language develop within and across sessions?}
\item \textbf{RQ4.} \textit{To what extent did conversational behavior predict same-day states and post-intervention outcomes?}
\end{itemize}

We find that whether the agent followed its prompt depended on the kind of rule. The rules it met were the checkable ones, like a cap on reply length, and it ignored the rules that described how it should behave, like being gently challenging, under both prompts. The conversations themselves looked more adaptive than we expected, with sessions ending in firmer plans rather than looser ones. The one behavior that tracked a worse outcome was the agent's habit of pressing participants to decide: the more demands to decide a participant received, the more doubt they reported at the end of the study. Because ours is an observational analysis, we treat that association as the most likely explanation the data offer. 

This paper contributes both a method for auditing conversation transcript and a set of finding about prompts. Instruction-following benchmarks test only the rules that can be checked automatically \cite{zhou2023ifeval}; we measure rules about agent disposition in a live deployment, with human coders and validated annotators doing the scoring, and find that those rules can fail in nearly every conversation without anyone noticing, because nothing in the output announces that they are being broken. The practical rule that follows is to write prompt rules you can verify from transcripts, and then verify them. The paper also contributes evidence about which agent behaviors matter in reflective conversations. The much-discussed flattery, measured at full strength in a real deployment, showed no link to any outcome, while repeated demands to decide, a behavior our codes make checkable in any transcript, was linked to greater career doubt at the end of the study. Together these narrow the explanations for the trial's result and yield a concrete design rule for reflection agents: budget decision demands and accept it when a participant declines to decide.

\section{Related Work}

\subsection{Reflection and journaling with language models}

HCI has studied technology for reflection for a long time, building on older accounts of how reflection works in professional and personal life \cite{schon1983}. Baumer \cite{baumer2015} described the different ways systems can support reflection, Fleck and Fitzpatrick \cite{fleck2010} mapped how deep reflection can go and Slov\'{a}k et al.\ \cite{slovak2017} argued that deep reflection needs more support than most systems provide. The practice these systems support also has a long history in clinical research: writing about one's experiences in a structured way has been studied as a therapeutic tool since the expressive writing experiments of the 1980s and 1990s \cite{pennebaker1997}. Conversational supports predate language models: relational agents maintained supportive relationships with users over weeks \cite{bickmore2005}, and Reflection Companion delivered scripted mini-dialogues about step data, showing that even canned follow-up questions prompt meaningful reflection \cite{kocielnik2018}.

Language models removed the need for scripts, and recent systems show the range of what became possible. MindfulDiary supported psychiatric patients' journaling through GPT-4 conversation in a four-week field study \cite{kim2024mindfuldiary}. DiaryMate examined sentence-level co-writing and observed users adopting the model's emotional framing in place of their own \cite{kim2024diarymate}. ExploreSelf generated adaptive questions for personal challenges \cite{song2025exploreself}, MindScape personalized prompts with behavioral sensing \cite{nepal2024mindscape} and CareCall documented what operating an LLM check-in service in public health requires \cite{jo2023carecall}. Reflection support of this kind is also moving beyond wellbeing apps into everyday tools, for example workplace agents that draw on activity data to prompt information workers to reflect on their work patterns \cite{nepal2025telemetry}. These studies show that such systems can be built and that people use them, and they offer real design insight. What they cannot show is whether conversation works better than cheaper formats, because most of them are short, small and have no comparison group. The trial we analyze \cite{soh2026tmb} is the exception: participants were randomly assigned, the comparison group received identical content and outcomes were measured at four points in time. Its result goes against the optimism of this literature, which is exactly why its conversations are worth reading closely.

\subsection{Measuring what deployed agents do}

Few studies have measured what an agent does turn by turn and connected that behavior to how users fare. Chiu et al.\ \cite{chiu2024} coded language models playing therapist and found behavior resembling low-quality human therapy, though their clients were simulated. Fang et al.\ \cite{fang2025} classified behaviors across 300,000 messages of a randomized chatbot-use study and related them to loneliness and dependence, which is the closest previous work to what we do here. Their setting was companionship, where people form ongoing relationships with chatbots \cite{skjuve2021}, rather than a structured intervention and their trial had no non-conversational arm.

Trials of conversational agents have rarely examined the conversations themselves. The earliest chatbot RCTs, such as the Woebot trial, reported outcomes and usage \cite{fitzpatrick2017}, systematic reviews of healthcare conversational agents describe the same pattern \cite{laranjo2018}, and the first randomized trial of a generative therapy chatbot reported symptom change and engagement, with conversations monitored for safety but not systematically analyzed \cite{heinz2025}. Nor is the pattern specific to therapy. A systematic review of 153 CHI papers engaging language models found the field studying closed models whose behavior can shift underneath a deployment, with authors repeatedly raising validity and reproducibility concerns \cite{pang2025llmification}, and auditing what a deployed agent actually did, from its own transcripts, is one concrete response to those concerns. Closest to our analysis of the agent's questions, Jacobsen et al.\ \cite{jacobsen2025probes} compared four kinds of follow-up questions in chatbot-run surveys and found that different kinds suit different research stages, though they measured the quality of the answers and the in-survey experience  rather than downstream effects on the respondent.

Research on sycophancy, the tendency of language models to tell users what they want to hear, gives one concrete reason to worry about unexamined agents. Sycophancy appears to be driven in part by the human preference judgments used to train models \cite{sharma2024sycophancy}, models affirm users far more than people do \cite{cheng2025elephant}, and in experiments, sycophantic responses make people more convinced they are right and less willing to repair conflicts \cite{cheng2026}. All of this evidence comes from the laboratory. Whether a deployed intervention agent actually behaves sycophantically, and whether that behavior harms the outcomes the intervention targets, has to our knowledge not been measured in the field, and the present study measures both. A similar gap holds for fidelity. Health intervention research treats verifying that the treatment was delivered as intended as basic practice \cite{bellg2004}, and instruction following in language models is explicitly trained \cite{ouyang2022} and constantly benchmarked \cite{zhou2023ifeval}. The benchmarks, however, restrict themselves by design to what can be scored automatically: IFEval, for example, tests only ``verifiable instructions'' and explicitly excludes instructions like writing in a funny tone, because there is no clear automatic standard for judging them \cite{zhou2023ifeval}. That restriction has left the other half of real system prompts, the instructions about manner, unmeasured. Prompt writers feel the gap from the other side: non-experts design prompts ad hoc and struggle to make instructions hold \cite{zamfirescu2023johnny}, with no way to see whether a given rule is working. This paper measures that half, using human coders and validated annotators as the scoring mechanism the benchmarks lack. Some pilot studies have begun to code chatbot transcripts against external therapy standards such as motivational interviewing \cite{suffoletto2025}, which is close in spirit, but the standard checked there is the therapy manual, not the instructions the agent was actually given, and to our knowledge no field study has audited a deployed agent's transcripts against its own system prompt.

Using language models to annotate text at scale is now common practice in computational social science \cite{ziems2024, gilardi2023, tornberg2023}. Because reliability varies by task, we validate our annotators against human coders for each field separately and use two model families rather than trusting any single model. What such an analysis should look for in the conversations is a separate question and we turn to it next.

\subsection{Which conversational behaviors to measure}
If transcripts are to explain outcomes, we need to know what to look for in them. Two literature point to behaviors worth measuring, and a third explains why the medium of the asking could matter. One is commitment language: counseling research shows that people's words carry signal about whether they are settling on a decision: how firmly clients state their commitments during motivational interviewing predicts whether they later act on them \cite{amrhein2003, miller2013}. This is why we rate how firmly participants word their decisions rather than just counting the decisions. The trial's outcome constructs come from research on identity development, in which people explore their options, settle on one and then come to feel that the choice is truly theirs \cite{marcia1966, luyckx2008, crocetti2008}; the outcomes that suffered in the trial's agent condition are the settling parts of that process.

Another is pressure to decide. Weighing options and acting on a choice are different mental modes, and the shift between them matters \cite{gollwitzer1990}. Pressure to reach a conclusion changes how people decide \cite{webster1994, kruglanski1996}, and self-determination theory expects choices made under outside pressure to be held less firmly than choices people arrive at freely \cite{ryan2000}. An agent that repeatedly asks someone to choose, endorses options while asking and asks again when the person hesitates would produce exactly this kind of pressure. In this paper, we check whether the trial's agent behaved this way.

A third tradition, with roots in media psychology and communication and long influential in HCI explains why the medium of a question could matter at all. Decades of experiments on computers as social actors show that people treat machines with conversational cues as if they were people, without meaning to and often while denying that they do it \cite{reeves1996, nass2000}. They are polite to computers, they manage the impression they make on computers, they disclose personal information to machines that ask for it \cite{moon2000}, and they answer a question differently when a computer asks it directly than when the same question sits in a form \cite{nass1999}. This matters for our setting because the two trial conditions asked the same questions through different media. A journaling form poses a question without being anyone: if the participant answers vaguely or skips ahead, nothing happens. A conversational agent poses the same question as a social counterpart. It is visibly waiting, it responds to what the participant says, and declining to answer it feels like declining a person. If being asked to make a decision carries a psychological cost, this tradition predicts the cost lands in the conversational condition, where the request comes from something participants treat socially. In this paper, our fourth research question tests that prediction.

\section{Study Context and Data}

We analyze data from a randomized trial of AI-guided versus self-guided career reflection, reported by Soh et al.~\cite{soh2026tmb}. The trial ran twice, first with students and then with a national online sample, randomized participants to reflect either in conversation with a GPT-4o agent or in a static journaling interface, and found that agent-condition participants ended the program less committed to their career plans. This section describes the trial's design, the two system prompts, the samples and the conversation corpus.

\subsection{Design and procedure}

\begin{figure*}
  \centering
  \includegraphics[width=\textwidth]{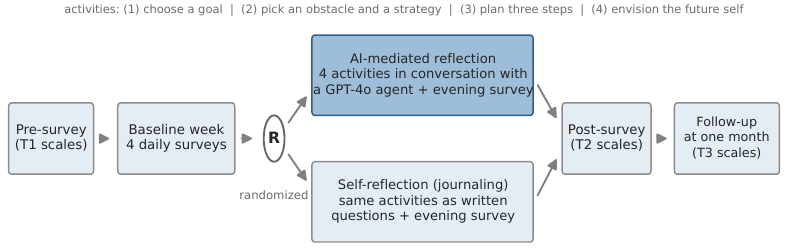}
  \caption{The trial's design. Both studies followed the same ten-day protocol, and the two conditions completed the same four activities in different formats.}
  \label{fig:design}
\end{figure*}

As mentioned earlier, the trial ran twice: first with community college and university students (Study 1) and then, after revisions, with a larger online sample recruited through Prolific (Study 2). Both studies followed the same ten-day protocol, shown in Figure~\ref{fig:design}. Participants completed a pre-survey, then four days of brief daily surveys that established each person's baseline week, then four days of career reflection activities each followed by the same daily survey, then a post-survey and a one-month follow-up. Randomization assigned participants to one of two conditions. Self-reflection participants answered a fixed sequence of open-ended questions per activity in a survey interface, ending each activity by writing a short summary of their main themes. AI-mediated participants completed the same activities in conversation with a GPT-4o agent \cite{openai2024gpt4o} on a platform built for the trial, which logged every message with a millisecond timestamp.

The four activities move from exploring toward planning. 
Activity 1 asks participants to identify the career goal that best fits their interests, values and skills. Activity 2 has them select one actionable obstacle and a strategy against it. Activity 3 turns the strategy into three concrete steps with timelines. Activity 4 has them envision their future self and their support system. Both conditions cover the same topics in the same order, so the conditions differ in medium rather than content.
The daily surveys asked participants to rate, on 0 to 100 sliders, five short statements about how they viewed their identity and future that day, drawn from a general daily identity measure \cite{hatano2020} (for example, ``Today, I had a clear view on my future''), along with their mood, anxiety and, on activity days, whether they had learned something about themselves. The pre-survey, post-survey and follow-up used two standard questionnaires about career identity, the 25-item DIDS \cite{luyckx2008} and the 30-item VISA \cite{porfeli2011}, which provided the trial's outcome measures.

\subsection{Two prompts, one model}

The trial team rewrote the agent's system prompt between the two studies, which gives the corpus a property we rely on throughout: the same model ran the same activities under two different sets of instructions. Table~\ref{tab:prompts} places the two prompts side by side on the rules that matter for our analysis, quoting each rule as written. Study 1's prompt casts the agent as a ``career reflection facilitator'' and controls it tightly. It tells the agent to ask one question at a time, collect at least fifteen responses per session, offer suggestions and insights, avoid overdoing positive feedback and end each session by writing a summary of the participant's reflections for them. Study 2's prompt, rewritten after Study 1 participants found the agent repetitive, casts it as a ``conversation partner'' and controls it loosely. It tells the agent to keep replies to one or two sentences, not to be ``overly agreeable or automatically validate everything the user says,'' to ask ``open-ended, clarifying, and gently challenging follow-up questions'' and to end by having the participant summarize their own themes. Both prompts also contain the same safety scripts: fixed wording for the agent to use if a participant mentions self-harm or asks to stop early, and a completion code the agent posts when an activity is finished. Appendix~\ref{app:prompts} reproduces the core instructions of both prompts, and the complete text of each is in supplementary materials. Because the sample and the prompt changed together between the studies, differences in agent behavior between them are descriptive: they could come from the prompt, the population or both.

\begin{table*}
\caption{The two system prompts, side by side, on the rules relevant to this paper. Quotes are verbatim from the deployed prompts. Appendix~\ref{app:prompts} gives the full core instructions.}
\label{tab:prompts}
\small
\renewcommand{\arraystretch}{1.25}
\begin{tabular}{@{}p{0.12\textwidth}p{0.41\textwidth}p{0.40\textwidth}@{}}
\toprule
 & \textbf{Study 1 prompt (``facilitator'')} & \textbf{Study 2 prompt (``conversation partner'')} \\
\midrule
Agent role & ``career reflection facilitator for emerging adults'' & ``conversation partner who helps the user reflect on their career path'' \\
Reply length & ``Avoid providing long responses---keep it concise, natural, and user-focused.'' & ``Keep your replies short (1--2 sentences)'' \\
Questions & ``Only ask one question at a time to not overwhelm the user.'' & ``Ask open-ended, clarifying, and gently challenging follow-up questions.'' \\
Praise & ``Do not overdo positive feedback---only validate when it feels necessary or appropriate'' & ``Do not be overly agreeable or automatically validate everything the user says.'' \\
Advice & ``Offer new suggestions or insights based on what the user shares.'' & not mentioned (not authorized) \\
Session length & ``Make sure user shares AT LEAST 15 responses.'' & no minimum; ``keep the conversation moving'' \\
Session ending & agent writes a summary of the participant's reflections & participant is asked to ``summarize the main themes'' themselves \\
\midrule
Safety scripts & \multicolumn{2}{p{0.81\textwidth}@{}}{identical in both studies: fixed wording for crisis disclosures and requests to stop early, and a completion code per activity} \\
\bottomrule
\end{tabular}
\renewcommand{\arraystretch}{1.0}
\end{table*}

\subsection{Samples and corpus}

Table~\ref{tab:corpus} summarizes the samples and the corpus. Study 1 ran in May and June 2025 with students from universities and community colleges in one metropolitan area. Study 2 ran in October and November 2025 with Prolific participants matched to census benchmarks on age, gender and ethnicity. Study 2 is larger and nationally representative. It is the study in which the trial's between-condition differences reached significance, and it is the study on whose transcripts we validated our annotators. It therefore serves as the primary study throughout, with Study 1 providing the second prompt and an exploratory replication sample.

Not every participant appears in every analysis, and the bottom rows of Table~\ref{tab:corpus} summarize the analysis samples. Of 222 participants randomized to the agent condition across both studies (84 in Study 1 and 138 in Study 2), 185 began at least one conversation and contribute transcripts, with the remainder mostly early dropouts who never started activity 1. The day-level models use the Study 2 sessions that linked to a same-day survey (98\% linked within one day), the person-level models use the participants with complete post-test scales and baseline covariates, and the one-month models use those with follow-up data.

\begin{table}
\caption{The two studies and the conversation corpus. Demographics describe each study's full sample as reported in the trial. Session length is computed from message timestamps and reported as the median because some sessions were paused and resumed across long gaps.}
\label{tab:corpus}
\small\setlength{\tabcolsep}{3.5pt}
\begin{tabular}{lrr}
\toprule
 & Study 1 & Study 2\\
\midrule
Recruitment & University sample& Prolific sample\\
Field period & May--Jun '25 & Oct--Nov '25\\
Age, mean (SD) & 20.8 (2.6) & 25.5 (2.7)\\
Women & 69.7\% & 46.6\%\\
\hdashline
Randomized to agent condition & 84 & 138\\
\quad with transcripts & 63 & 122\\
Sessions & 230 & 457\\
Sessions per participant & 3.7 & 3.7\\
\hdashline
Agent turns & 2,382 & 6,935\\
Participant turns & 2,122 & 6,491\\
Participant turns per session & 9.2 & 14.2\\
Words per agent turn & 58.6 & 40.6\\
Words per participant turn & 20.3 & 27.0\\
Words per participant, total & 684 & 1,434\\
Median session length & 14.2 min & 20.1 min\\
\hdashline
Sessions in day-level analysis & --- & 448\\
Participants in person-level analysis & --- & 107\\
\quad with one-month follow-up & --- & 89\\
\bottomrule
\end{tabular}
\end{table}

The conversations are substantial. A typical Study 2 participant exchanged over one hundred messages with the agent and wrote about 1,400 words of their own. Journaling participants wrote a nearly identical amount across the same activities (about 1,440 words on average in the trial's response data), so the two conditions produced comparable volumes of participant writing.

\subsection{Privacy and ethics}
\label{sec:ethics}
The trial was approved by the institutional review board at Stanford University. We scanned all 17,930 turns for direct identifiers before analysis, finding none. Because free text still carries indirect clues such as named schools, employers and first names of friends and family, we quote conversation turns only as paraphrases \cite{swain2020}, we quote post-survey feedback verbatim only after screening it for identifiers, and verbatim transcripts do not leave the research team. For annotation, we sent the deidentified transcript text, without participant identifiers, to the annotator models through OpenRouter, an API gateway whose policy is not to store prompts. The model providers' API terms likewise exclude training on submitted data and limit any retention to short abuse-monitoring windows, typically up to 30 days.

\section{Measuring Conversational Behavior}

Our measurement has three parts: a coding scheme, evidence that two humans can apply it consistently, and tests showing that the language model annotators who applied it at scale match the humans. We share the full codebook, with definitions, decision rules and worked examples, in supplementary materials. Table~\ref{tab:examples} illustrates the agent-side categories, and Table~\ref{tab:excerpt} shows the codes applied to a real exchange.

\subsection{Coding scheme}
 
The agent and the participant do different jobs in these conversations, so their turns need different codes. The agent's job is to act: it asks, praises, advises and presses, often several of these at once, so we code which actions each agent turn contains. The participant's job is to hold a position about their own career, so each participant turn gets one category for what that position is, along with ratings for how firmly it is held and whether it contains an insight. We coded each agent turn for the presence of nine moves. An \textit{open question} invites elaboration. A \textit{closed question} calls for a yes, a no or a choice among stated options. \textit{Restatement or synthesis} mirrors or interprets what the participant said. \textit{Validation} covers praise, approval and reassurance, including brief markers such as ``Great---.'' \textit{Challenge} introduces a doubt, tension or counterpoint the participant had not raised. \textit{Advice} proposes actions, resources or information the participant did not supply. \textit{Process talk} manages the task. A \textit{closure move} asks the participant to settle on one goal, obstacle, strategy or step to carry forward; because these turns demand a decision, we also call them decision demands. \textit{Question stacking} flags turns demanding two or more distinct answers. Agent turns usually do several things at once, averaging 2.8 moves per turn, so a turn can receive several codes. The categories come from the response categories counseling researchers use to code what therapists do \cite{hill2014}, which have been applied to language models before \cite{chiu2024}. We added closure moves and question stacking because the trial's activities are built around decision points.

We gave each participant turn one primary category: exploring, committing, describing a barrier or distress, revising an earlier position, or a minimal response. Coders also flagged \textit{insight}, a conclusion about oneself drawn from experience, and rated \textit{certainty} from 1 to 5, skipping turns that took no position about the participant's career. These categories are adapted from how motivational interviewing researchers code clients' ``change talk'' \cite{miller2013, amrhein2003}, adjusted to the career constructs the trial measured \cite{luyckx2008}.

Before the results, the certainty and challenge codes need a word of explanation. Certainty is rated from the wording of the turn, with examples fixed at each scale point: a 1 is a stance dominated by hedging (``maybe something with art, I really don't know''), a 3 is a commitment with an explicit qualifier (``I guess accounting, if the classes go okay'') and a 5 is an unqualified statement of intent (``I'm going to apply to the nursing program in January''), with 2 and 4 for stances leaning each way. Coders rate the words in front of them, not their overall impression of the person, which keeps ratings comparable across turns and speakers. Challenge, in turn, requires that the doubt come from the agent. If a participant voices a worry and the agent asks them to say more, that is a follow-up question. If the agent questions the feasibility of a plan the participant is happy with, that is a challenge. This distinction turned out to be the hardest judgment in the codebook: it is why challenge was the code our two human coders agreed on least, and, as we show below, the code where the choice of annotator model mattered most.

\begin{table}
\caption{Agent move categories with paraphrased examples. Most turns combine two or three moves.}
\label{tab:examples}
\small
\begin{tabular}{p{1.85cm}p{5.5cm}}
\toprule
Move & Paraphrased example\\
\midrule
Open question & ``What was it about theater that pulled you in as a kid?''\\
\hdashline
Closed question & ``Does that career still feel right, or would you rather focus on a different one?''\\
\hdashline
Restate / synthesize & ``It sounds like steady effort and support from your family are what keep you going.''\\
\hdashline
Validation & ``That's a solid plan.'' / ``Great---that clarity helps.''\\
\hdashline
Challenge & ``If your posts don't get the attention you hope for, how would you adapt?''\\
\hdashline
Advice & ``You could look into free courses or reach out to someone already in the field.''\\
\hdashline
Process & ``Let's move to your second step.''\\
\hdashline
Closure move & ``Of the paths you've named, which one feels like the best fit to carry forward?''\\
\hdashline
Stacking & ``What exactly will you do, when will you finish it, and what might get in the way?''\\
\bottomrule
\end{tabular}
\end{table}

\begin{table}
\caption{A repeated decision demand, with our codes. The participant gives an integrative answer, combining the options rather than picking one; the agent re-poses the demand and the eventual commitment is hedged. Sequences like this followed 15.3\% of answered closure moves.}
\label{tab:excerpt}
\small
\begin{tabular}{p{1.6cm}p{5.6cm}}
\toprule
Codes & Exchange (paraphrased)\\
\midrule
closure move, validation & \textit{Agent:} Between reaching out for advice and applying broadly, which do you think would work best for landing a better-paying job?\\
\hdashline
no commitment (integrative) & \textit{Participant:} Honestly, a combination of both.\\
\hdashline
closure move repeated, validation & \textit{Agent:} That makes sense, combining them could open more doors. Which one feels like the right next step for you personally, the one you would realistically try first?\\
\hdashline
commitment, certainty 3 & \textit{Participant:} I guess reaching out first, maybe next week.\\
\bottomrule
\end{tabular}
\end{table}

\subsection{Reliability between human coders}

We developed the scheme in two rounds. In a calibration round, two coders coded the same 40 Study 2 turns, compared results and revised the codebook where they had disagreed. The two coders then independently coded a fresh sample of 100 agent turns and 100 participant turns, drawn from across activities and session positions. Each turn was shown with the two turns that came before it, so coders had the local context. Table~\ref{tab:reliability} reports how well the two coders agreed. Ten of twelve fields reached Cohen's $\kappa \geq .65$, which conventional benchmarks describe as substantial agreement \cite{landis1977}, and certainty reached a weighted $\kappa$ of .79, with 99\% of rating pairs within one point of each other. The two exceptions, challenge and insight, are the scheme's rarest codes, and $\kappa$ has a known problem with rare codes: even coders who agree on nearly every turn get a low score, because agreeing that a rare thing is absent earns little credit \cite{byrt1993}. For those two fields we therefore also report a prevalence-adjusted version of $\kappa$ (PABAK), which corrects for rarity. The doubly coded set is about one percent of the corpus, but the precision of a reliability estimate depends on the number of items coded rather than their share of the corpus and reliability in HCI practice is conventionally assessed on a coded subset of this size rather than on the full corpus \cite{mcdonald2019}. The corpus itself has a separate full-coverage safeguard: both annotator models labeled every turn, and their corpus-wide agreement, reported below, checks all 17,930 turns rather than a sample.
\begin{table}
\caption{Reliability and annotator validation. Human agreement is between two independent coders (100 turns per instrument). Annotator agreement is against the consensus gold standard (items where the coders agreed), on held-out items never shown as prompt examples. A = gpt-5.6-luna-pro. B = claude-sonnet-5. Both run at temperature 0 with a frozen prompt. PABAK is prevalence-adjusted $\kappa$. Certainty is evaluated as mean absolute error (MAE) against the coders' mean on the 1--5 scale: an MAE of .32 means the annotator's rating differs from the coders' average by about a third of a scale point.}
\label{tab:reliability}
\small\setlength{\tabcolsep}{3.5pt}
\begin{tabular}{lccc}
\toprule
Field & Human $\kappa$ (PABAK) & Annotator & Annot.\ $\kappa$\\
\midrule
Open question & .90 (.92) & A & .87\\
Closed question & .82 (.84) & A & .94\\
Restate / synthesize & .72 (.78) & A & .97\\
Validation & .72 (.72) & A & .95\\
Challenge & .34 (.86) & B & .85\\
Advice & .65 (.92) & A & .85\\
Process & .74 (.74) & A & .97\\
Closure move & .80 (.84) & A & .85\\
Question stacking & .72 (.74) & A & .94\\
\hdashline
Participant category & .77 & A & .81\\
Insight & .54 (.76) & B & .50\\
Certainty (weighted) & .79 & A & MAE .32\\
\bottomrule
\end{tabular}
\end{table}

\subsection{Validated language model annotators}

Hand-coding all 17,930 turns was not practical, so language models did the labeling, but only after we tested them against the human coders. The test worked as follows. From the 200 doubly coded turns, we kept the items where both humans agreed as the answer key (86\% to 96\% of items per field; for certainty, the answer key is the average of the two ratings). We then gave each of four candidate models the codebook plus a small set of worked examples in its prompt (a few-shot setup; Appendix~\ref{app:annotatorprompt} reproduces the prompt) and had it code the remaining items without seeing the answers. The selection rule, fixed before we looked at any scores, was to deploy, for each field, a model whose agreement with the human answer key came within .05 of the agreement between the two humans, preferring to keep one mid-priced model across all the fields where it qualified rather than mixing several. Appendix~\ref{app:annotators} reports every model's scores. A mid-priced model (gpt-5.6-luna-pro) passed this bar for ten of the twelve fields. It failed the two rare codes, challenge and insight, which both require judging whether an idea came from the participant or from the agent. A more capable model (claude-sonnet-5) qualified for both, agreeing with the human answer key on challenge well above the human coders' own agreement. The two selected models then labeled every turn in both studies, using identical frozen prompts at temperature 0. As a further check, we had both models label the entire corpus and compared them: they agreed well on all ten shared fields ($\kappa$ .70 to .89 in Study 2 and .64 to .86 in Study 1), so no finding in this paper depends on which model produced the labels, apart from the two fields only one model was qualified to code.

Before trusting the participant-side measures, we ran one more check. Our certainty and insight codes assume that participants' words are their own, but chat participants sometimes answer by echoing the question back, and a participant who copies the agent's phrasing would inherit the agent's confidence rather than express their own. We therefore measured, for every participant turn, how much of its wording overlapped with the agent's two preceding turns. Only two turns out of 6,491 overlapped more than 20\%, so copying was essentially absent and the participant-side measures most likely reflect participants' own language.

\subsection{Analysis}
\label{sec:analytic}
To limit the room for cherry-picking, we fixed several decisions before looking at the relevant data. We fixed the annotator selection rule described above. We decided that every family of significance tests would be corrected for multiple comparisons with the Benjamini-Hochberg procedure \cite{benjamini1995}. We decided in advance how to read the RQ4 results: if specific agent behaviors reliably predicted outcomes, that would point to those behaviors as the explanation for the trial's result, and if nothing did, that would point instead to something all the agent conversations shared. Finally, for the post-survey analyses in Section~\ref{sec:perceptions}, we wrote down two predictions before opening those data: participants who received more decision demands would list more career options still open at post-test, and they would report more difficulty from conflicting advice from other people. We committed to reporting both tests however they came out, and everything else in that section is labeled exploratory or descriptive.

Each research question then gets the simplest analysis that answers it. RQ1 and RQ2 are descriptive: we compare the agent's behavior rates against what its own prompt asked for. For RQ3, we track the certainty ratings and participant categories across positions in the session, compare decisions the agent asked for with decisions participants volunteered, and compare the closing reflections of the two conditions using a simple hedge-word count applied identically to both, since the validated certainty measure exists only for conversation turns. For RQ4, we ask two versions of the outcome question. At the day level, we test whether a participant's conversation on a given day predicted their survey that evening, using mixed models across 448 session-days that compare each participant with their own typical day (with activity fixed effects, baseline-week covariates and random intercepts). At the person level, we test whether a participant's overall conversation profile predicted their post-test scores, using regressions across the 107 participants with complete data (controlling each outcome's baseline value, age and gender, with robust standard errors).

\section{Results}

\subsection{Fidelity to the system prompt (RQ1)}
\label{sec:rq1}
\begin{figure*}
  \centering
  \includegraphics[width=\textwidth]{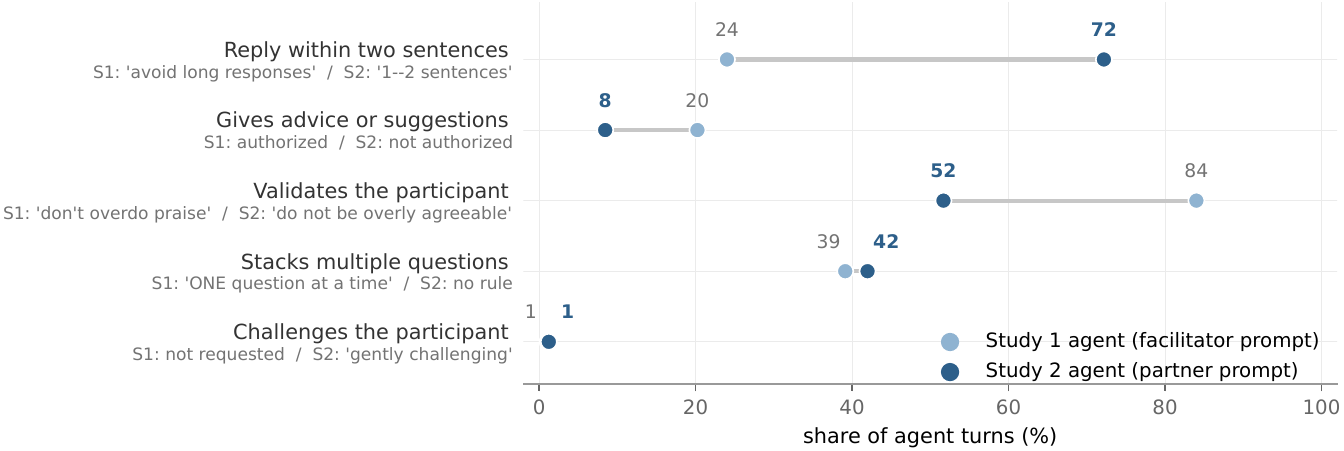}
  \caption{Agent behavior compared with what each study's prompt asked for. Rules that can be checked from the output, like the reply length cap and whether advice was allowed, changed the agent's behavior when they changed between studies. Rules about manner did not: the agent challenged participants in about one percent of turns whether the prompt asked for challenge (Study 2) or never mentioned it (Study 1).}
  \label{fig:spec}
\end{figure*}

Did the agent do what its prompt told it to do? The answer depends on the kind of rule, and Figure~\ref{fig:spec} compares what each study's prompt asked for with what the agent did. The prompts contain two kinds of rules. Countable rules set targets that anyone can check from the output, such as the two-sentence cap on replies. Dispositional rules describe how the agent should behave or its manner, such as not being overly agreeable. The agent largely followed the countable rules: 72\% of Study 2 replies fit the two-sentence cap, scripted endings appeared in 98\% of sessions, and the pause and crisis protocols fired with their scripted wording in the handful of turns that called for them. It did not follow the dispositional rules, and the deviations all pointed in the same direction. Told not to be overly agreeable, it validated in 51.7\% of turns. Our validation code is deliberately broad, counting brief openers like ``Great---'' as well as substantive praise, and in the doubly coded sample the coders flagged about 40\% of validation codes as such brief markers. Even counting only praise with real content, then, the agent validated in roughly a third of its turns, against an instruction telling it not to. Told to summarize occasionally, it restated or synthesized in 84\% of turns. Told to ask gently challenging follow-ups, it challenged in 1.2\%, and over half of participants finished all four activities without a single challenging turn. Neither the validation figure nor the challenge figure matched our expectations, so we checked both against every instrument we had before accepting them. In short, every rule the agent followed was one it could be graded on, and no rule about how to behave was followed. The reverse does not hold, as the next comparison shows: being gradable did not guarantee a rule was followed.

Because the two studies gave the same model different instructions (Table~\ref{tab:prompts}), we can also ask which behaviors moved when the instructions changed, with the caution from Section 3.2 that the population changed as well. Table~\ref{tab:twoagents} shows both agents side by side, and the same split appears. Behaviors covered by countable rules moved with the rules. Study 1's vague ``avoid long responses'' produced 3.8 sentences per turn, while Study 2's explicit two-sentence cap produced 2.3. Advice was allowed in Study 1 and not mentioned in Study 2, and it fell from 20\% of turns to 8\%. Validation fell from 84\% under Study 1's mild caution to 52\% under Study 2's stronger wording, so a firmer instruction may have helped, though the population changed along with the prompt, and either way the agent still praised in half its turns. Other rules changed nothing. Study 1 told the agent to ask one question at a time, yet about 40\% of turns stacked multiple questions in both studies. Study 1 required at least fifteen participant responses per session, and only 7\% of its sessions reached that number. And challenge sat at about 1\% in both studies, even though only Study 2 asked for it. That last comparison is the cleanest evidence in the paper: an entire instruction, present in one prompt and absent from the other, changed nothing.
That said, the challenge numbers rest on the code our human coders agreed on least ($\kappa=.34$), and in any case the finding here is simply how rare the behavior was. We can say that this model, under these two prompts, would not challenge. Whether any prompt could make current models challenge is a bigger question, and our data cannot answer it. Decision demands also differed between the studies. Closure moves per participant doubled from 4.8 to 10.1 (Table~\ref{tab:twoagents}), consistent with the activities' built-in decision points combined with Study 2's shorter, faster-moving turns.

\begin{table}
\caption{Behavioral profile of the two deployed agents. Both ran GPT-4o on the same four activities under different prompts. Countable rules (length, advice) moved with their instructions; dispositional rules moved partly (validation) or not at all (challenge), and neither reached compliance.}
\label{tab:twoagents}
\small
\begin{tabular}{lrr}
\toprule
 & Study 1 & Study 2\\
\midrule
Words per turn & 58.6 & 40.6\\
Sentences per turn & 3.8 & 2.3\\
Turns within two sentences & 24\% & 72\%\\
Question stacking & 39\% & 42\%\\
Validation & 84\% & 52\%\\
Challenge & 1.1\% & 1.2\%\\
Restate / synthesize & 89\% & 84\%\\
Open questions & 78\% & 79\%\\
Advice & 20\% & 8\%\\
Closure moves per participant & 4.8 & 10.1\\
Ending protocol executed & 97\% & 98\%\\
Participant certainty, mean & 4.00 & 4.00\\
Participant insight rate & 6.5\% & 8.6\%\\
\bottomrule
\end{tabular}
\end{table}

The agent also drifted more as sessions wore on and as the task grew heavier. Early in a session, 82\% of replies fit the length cap; past the fifteenth turn, only 62\% did. The worst stretch was activity 3, whose script asks participants to produce three concrete steps with four details each. Under that load, the agent often bundled several questions into one turn and asked for a commitment at the same time, which in practice meant reading the journaling condition's questionnaire aloud inside the chat. There was also one failure we specifically looked for and never found. The agent almost never answered the reflective questions for the participant: in the entire corpus, we found only two turns where it slipped into long, unprompted advice. So the agent bent its rules where the task was heaviest, but the reflective work stayed with the participant throughout.

\subsection{The agent's conversational moves (RQ2)}
\label{sec:rq2}
The previous section asked whether the agent followed its rules. This section asks what it actually did with its turns. Unless marked otherwise, statistics from here through Section 5.4 describe Study 2, the primary study. The agent's most common turn combined restatement, validation and an open question (30\% of turns), followed by restatement with an open question (27\%), a loop that mirrors, affirms and probes. Cross-session memory worked as designed, with 75\% of session openers in activities 2 through 4 recapping the participant's previous decision. Between participants, however, behavior varied considerably. One person's agent praised them in a quarter of its turns and another's in three quarters (the range ran 24\% to 73\%). Challenge shows the same unevenness in starker form. Challenges made up only about 1\% of turns overall, but those few turns were spread unevenly across people: 45\% of participants received at least one challenge somewhere in their four sessions, and the other 55\% never received any. Two people could complete the same program and, in this respect, receive different treatments.

Study 2's prompt told the agent not to ``automatically validate everything the user says'' (Table~\ref{tab:prompts}), and that wording invites a specific test: was the praise automatic, or did it respond to what the participant said? To find out, we computed the validation rate separately by what the participant had just done, and the rate barely moved. Whether the participant was exploring options, describing a barrier or a hard day, revising a plan or typing a minimal answer under ten words, praise followed about half the time (between 48\% and 57\%), and it peaked at 67\% after the participant committed to something. Participants who sounded unsure drew slightly more praise than participants who sounded confident, and after turns where the participant took a position, the agent affirmed 42 times more often than it challenged. The praise, in other words, responded to the presence of a participant rather than to the content of the answer. We call this pattern indiscriminate affirmation, and participants appear to have registered it as comfort rather than judgment: asked at post-test how the study changed how they think about themselves, one wrote that ``it's just reaffirmed things I already knew,'' and another that they felt ``more confident and self affirmed on my desired career path now that I've thought about it and discussed it.'' The pattern matches what Cheng et al.\ call the emotional-validation side of social sycophancy \cite{cheng2025elephant}. It is different from sycophancy in the stricter sense of agreeing with whatever position the user takes regardless of its merits \cite{cheng2026, sharma2024sycophancy}. Our codes cannot measure that stricter kind, because no annotator can score the objective merit of a stranger's career choice.

Two more patterns come back later in the paper. Nearly half of closure moves (45\%) carried validation in the same turn, endorsing an option while asking the participant to select it, so the agent's decision points were rarely neutral. One participant described this experience directly, writing that the study ``didn't really teach me new methods or provide any resources. I think it just focused on helping me make a decision and encouraging that I take action on that decision.'' Participant language, meanwhile, followed the intervention's intended arc, exploring in activity 1 (77\% of turns) and committing in activity 3 (57\%). Insight, however, concentrated almost entirely in the autobiographical first activity, at 14.6\% of turns there against 2.8\% during planning. Whatever self-insight the design produced, it produced almost entirely on the first day.

\subsection{Commitment language within and across sessions (RQ3)}
\label{sec:rq3}

\begin{figure}
  \centering
  \includegraphics[width=0.75\linewidth]{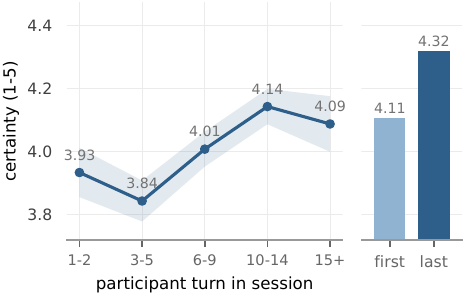}
  \caption{Participant certainty within Study 2 sessions. Left: mean certainty by turn position with a 95\% confidence band. Right: certainty of each session's first and last participant turns.}
  \label{fig:certainty}
\end{figure}

This section follows how firmly participants held their plans as their conversations unfolded. Going in, we expected the conversations to open decisions up without closing them. The trial had reported more tentative language in the agent condition \cite{soh2026tmb}, our own early analysis pointed the same way, and we built the certainty measure partly to catch that failure. The measure came back showing the opposite (Figure~\ref{fig:certainty}). Within a session, participants' certainty dipped a little in the middle and rose toward the end, and sessions typically ended with participants sounding more sure than when they started (first turns averaged 4.11 on the five-point scale, last turns 4.32; 36\% of sessions ended higher and 24\% ended lower).

Our turn-level ratings also help resolve a puzzle the trial itself flagged. In its person-level language analysis, agent participants' writing scored higher on insight ($\beta=.14$), higher on tentativeness ($\beta=.19$) and, at the same time, higher on certitude ($\beta=.19$), and the trial noted that tentativeness and certitude do not usually go together and could only speculate about why they co-occurred \cite{soh2026tmb}. The timing in our data offers an answer. A conversation that wavers in the middle and settles at the end contains both hedging words and confident words, so a count over the whole conversation picks up both, whereas turn-by-turn ratings show when each appears. The two measures are different instruments, so we offer this as a reading that fits both results rather than as proof. Some participants described the same arc themselves: one wrote that the conversations made them ``question deeper why I'd like to commit to my goals,'' which in turn ``made me realize I really would like to stick to them.''

We also compared decisions the agent asked for with decisions participants offered on their own, and found no difference in how firmly they were worded: demanded decisions averaged 4.31 in certainty and spontaneous ones 4.35, with weakly worded commitments equally rare in both (10\% against 9\%). Nor did participants often walk decisions back. Only 115 of 6,491 participant turns revised an earlier position, most of those came in the first activity, and most came without the agent having just asked for a decision, which looks like people correcting their own first drafts rather than being shaken loose. Finally, the short written reflections that closed each activity contained about the same amount of hedging in both conditions (0.44 against 0.55 hedging words per hundred words, not a significant difference).

The one place things went differently was when a participant declined to decide, and Table~\ref{tab:excerpt}, which we used earlier to illustrate the codes, shows a real example of this sequence. Of the 1,145 decision demands that got an answer, 15.3\% got an answer that was not a decision: sometimes the participant combined the offered options instead of picking one, and sometimes they simply said they did not know. In these cases, the agent almost always asked again. After that second ask, just over half of these exchanges (55\%) ended in a commitment. These commitments were the weakest in the corpus, averaging 3.82 in certainty, about half a point below every other kind. One exchange in nine ended with the participant describing obstacles or frustration instead of deciding. This is where the corpus's weakly held commitments come from. Participants noticed this behavior on their own: one wrote at post-test that ``the bot sometimes would just rephrase their questions rather than asking new things,'' a plain description of the 175 repeated demands our codes counted. So the conversations closed well, with one exception: when a participant declined to decide, the agent asked again, and what it got back was weaker.

\subsection{Conversational behavior and outcomes (RQ4)}

The final question is whether any of this behavior mattered for how participants fared. We looked first at single days: did a participant's conversation on a given day predict how they felt that evening? We tested every combination of eight conversation features and six evening survey measures, 48 tests in all, and none of them showed a reliable link, even before correcting for multiple testing (Figure~\ref{fig:null}; full numbers in Appendix~\ref{app:daylevel}). However, one link looked real at first, and it is worth explaining why it was not. In simple correlations, participants who received more praise also reported more career doubt. But that link reflected who those participants were, not what the praise did to them: people who were doubtful at baseline drew more praise from the agent, and once we accounted for baseline doubt, the link disappeared. Praise exposure also varied a lot from person to person, from 24\% of turns to 73\%, so if praise were causing day-to-day harm anywhere in that range, these tests were positioned to detect it, and they did not. Day to day, how a participant's agent behaved made no detectable difference to how their evening went.

\begin{figure}
  \centering
  \includegraphics[width=0.75\linewidth]{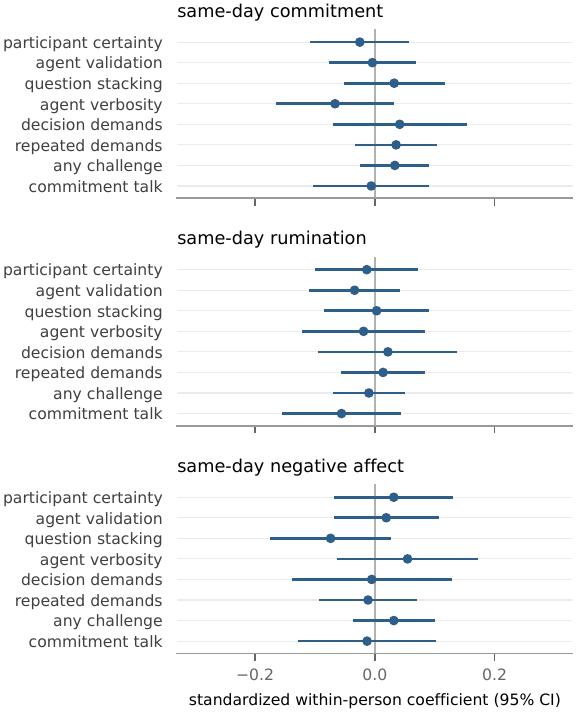}
  \caption{Each day's conversation features and that evening's self-reports. Every dot is one estimate, with its 95\% confidence interval, from mixed models that compare each participant with their own typical day. All intervals cross zero: no feature of a day's conversation predicted how the participant felt that evening.}
  \label{fig:null}
\end{figure}

\begin{figure}
  \centering
  \includegraphics[width=0.75\linewidth]{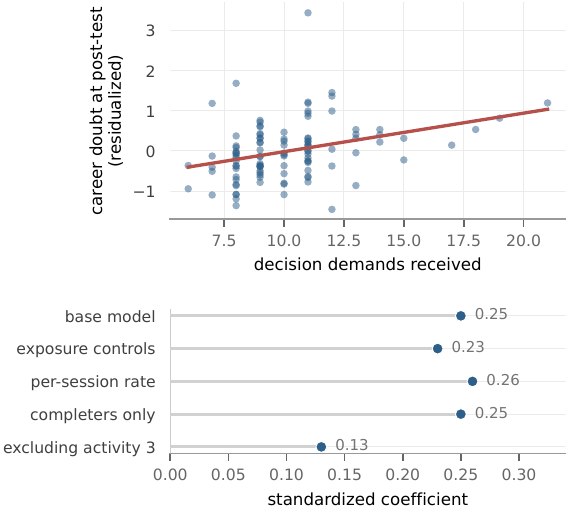}
  \caption{Decision demands and career doubt in Study 2. Top: each point is one participant, showing their post-test career doubt (adjusted for baseline doubt, age and gender) against the number of decision demands their agent made; the line shows the trend. Bottom: the same estimate recomputed under alternative versions of the analysis. Participants who received more demands reported more doubt ($\beta=.25$), and the estimate stays similar across the checks, weakening only when activity 3, the heaviest-demand day, is excluded.}
  \label{fig:doubt}
\end{figure}

We then looked at totals: did participants who experienced more of a behavior over the four days end the study differently? Out of 32 tests, one association held up. Participants whom the agent asked to decide more often reported more doubt about their plans at post-test ($\beta=.25$, 95\% CI [.14, .36], adjusted $p<.001$), over and above their baseline doubt, age and gender (Figure~\ref{fig:doubt}). This is not simply a matter of some participants talking more, because the association holds when we control for sessions completed, time on task and messages written, and it holds when we use demands per session instead of the total. It also holds when we drop any single activity from the count, though it weakens when we drop the planning activity, the day with the most demands ($\beta=.13$, $p=.04$). Appendix~\ref{app:personlevel} reports all of the checks. The association is robust, but it is still a correlation, and it is the only one that survived correction: nothing else did, and nothing at all predicted the one-month follow-up. If anything, more demands went with slightly more commitment making before correction ($\beta=.13$), which fits the picture from RQ3 of demands producing commitments in the moment while doubt builds by the end. Across the whole grid, one behavior, the demand to decide, reliably tracked one outcome, doubt about the decision.

In Study 1 the association pointed the same way but was not significant ($\beta=.10$, $p=.40$, n=51). Study 1 was smaller and its agent made fewer than half as many decision demands per participant, so we treat this as consistent with Study 2 rather than as a replication of it.

However, the demands were not randomly assigned. The agent asked again whenever a participant failed to choose, so a participant's own indecision raises both the number of demands they received and, plausibly, their later doubt. Controlling for baseline doubt helps, and the repeated-demand pattern in Section~\ref{sec:rq3} shows the pressure operating in the harmful direction turn by turn. Even so, we present decision pressure as the best-supported explanation in these data, not as a proven cause.

\subsection{How participants described the experience}
\label{sec:perceptions}
Finally, as a further check on RQ4, the post-study survey lets us compare these findings with what participants reported about themselves, following the plan in Section~\ref{sec:analytic}: two predictions written down in advance, everything else exploratory, agent condition only (n=107). We had predicted that participants who were pressed to decide more often would keep more career options open at post-test and would feel more torn by other people's opinions (measured with items from the career decision difficulties taxonomy \cite{gati1996}). Neither happened (Table~\ref{tab:confirmatory}). Decision pressure was linked to how securely participants held their plans, and to nothing else nearby, which narrows the doubt finding.

\begin{table}
\caption{The two pre-specified tests: decision demands received predicting post-test perceptions. Standardized coefficients with robust errors, controlling age, gender and, for external conflict, the same items at baseline. Directions were predicted before the data were opened. Neither test reached significance.}
\label{tab:confirmatory}
\small\setlength{\tabcolsep}{3.5pt}
\begin{tabular}{lccc}
\toprule
Outcome at post-test & Pred. & $\beta$ [95\% CI] & $p$ \\
\midrule
Career options listed (M=2.6, SD=1.5) & $+$ & .01 [$-$.19, .21] & .94 \\
External-conflict difficulty & $+$ & .12 [$-$.02, .26] & .10 \\
\bottomrule
\end{tabular}
\end{table}

The remaining checks showed nothing unusual. Participants found the activities engaging (91\% at least somewhat) more than enjoyable (75\% at least somewhat), and how much praise a participant received predicted neither rating, nor their self-reported reflection habits (all $|\beta| \le .11$). The post-survey also asked participants, in their own words, how the study had changed the way they think about themselves. We labeled these 106 written answers with the same validated annotators we used for the conversations. About two thirds of the answers described no change at all: participants wrote that the study had confirmed what they already believed about themselves and their plans. This is consistent with what the praise appeared to do inside the conversations (Section~\ref{sec:rq2}), reassuring people about the path they were already on rather than moving them to a new one.

\section{Discussion}

Our results answer two questions at once. About prompts, they show that every rule a deployed agent followed was one it could be graded on, and that no rule about how to behave was followed, under two different prompts. About the trial, they clear the agent of misbehavior and of harmful flattery and leave decision pressure as the most likely explanation for the added career doubt the trial observed.

\subsection{Implications for designing and auditing system prompts}
\label{sec:promptimplications}
The central lesson for prompt writers is that fidelity and effectiveness are different things, and both need checking. Where the agent followed its prompt, it followed the checkable parts, and the one behavior linked to harm, the demand to decide, was itself scripted: each activity directs the agent to get the participant to settle on one goal, obstacle, strategy or set of steps. When an agent-based intervention disappoints, the field's reflex is to blame the prompt. Here the prompt worked where it could work at all, and the outcome still went the wrong way.

Someone might say we already knew part of this. Anyone who has written a system prompt has watched a model ignore an instruction about tone. But there is a difference between knowing that a rule sometimes slips and knowing that it does almost nothing. The agent challenged in 1.2\% of its turns, and more than half of participants never received a single challenging turn. The two prompts show something belief alone cannot: Study 1's prompt never asked for challenge, Study 2's asked for it directly, and the rate was the same either way. Prompts for interventions like this one are written by domain experts, and the wording here reads the way you would brief a human facilitator: do not be overly agreeable, ask gently challenging questions. That is the natural way to describe how a person should behave, it is how non-experts approach prompt writing generally \cite{zamfirescu2023johnny}, and it is part of why this failure is easy to miss.

The design response follows directly: write prompt rules you can check, and check them. Every rule that changed the agent's behavior in this corpus could be graded from the output, and no rule about manner changed it. Table~\ref{tab:rewrite} shows what rewriting looks like for the prompt we studied. Instead of ``do not be overly agreeable,'' the rule becomes: at most one praising phrase per reply, and no praise in a reply that asks for a decision. We cannot promise that rewritten rules will work better, but when they fail, someone will notice. A rate that drifts can be watched and corrected, whereas ``be gently challenging'' failed ninety-nine times out of a hundred and nobody saw it. Some behavior may not be reachable by any wording at all, and challenge looks like that in our data. There the fix is a different design rather than better wording: give a second model the single job of raising one counterpoint per session, then count how many were delivered and report the number.

\begin{table}
\caption{Dispositional prompt rules from the trial, the behavior observed under them, and verifiable rewrites this corpus motivates. The length rule is included as the counterexample: stated as a countable quantity, it largely worked.}
\label{tab:rewrite}
\small
\begin{tabular}{p{0.27\linewidth}p{0.17\linewidth}p{0.44\linewidth}}
\toprule
Rule as written & Observed & Verifiable rewrite \\
\midrule
``do not be overly agreeable,'' do not ``automatically validate'' & validation in 51.7\% of turns & at most one affirming clause per turn; no validation in the same turn as a decision demand \\
\addlinespace
ask ``gently challenging'' follow-up questions & challenge in 1.2\% of turns & a separate critic pass generates one counterpoint per session; delivered challenges are logged and reported \\
\addlinespace
keep replies to 1--2 sentences & 72\% within cap (92\% within three) & keep as written; already verifiable per turn \\
\bottomrule
\end{tabular}
\end{table}

A fair question is whether any of this survives the next model release. The audit itself does not depend on the model: whatever is deployed, the only way to know whether a rule about manner is being followed is to measure it, and our codebook, annotators and selection rule can be reused as they stand. The split between the two kinds of rules also has a structural reason to persist, because both training and benchmarking reward what can be scored \cite{ouyang2022, zhou2023ifeval}, and what can be scored is the countable half. The decision-pressure mechanism belongs to the conversational medium and to people rather than to GPT-4o \cite{reeves1996}. And where models do change, that is an argument for the audit rather than against it: these findings are a dated baseline, and re-running the measurement is how a team would learn whether a new model has changed the picture.

\subsection{Implications for AI-guided reflection}

For reflection programs, we can start by ruling out some explanations for the trial's result. The agent did not malfunction: it ran its safety and ending scripts, kept its replies short and left the reflective work to the participant. The conversations did not ramble: sessions ended with participants sounding more decided than when they started, consistent with the reading of the trial's language findings in Section~\ref{sec:rq3}. And the flattery did not do the damage: praise exposure varied threefold across participants, and the heavily flattered did no worse than the lightly flattered, on any outcome, at any point. What we measured here is praise and emotional validation, what recent work calls social sycophancy \cite{cheng2025elephant}. We could not measure the stricter kind, agreeing with a user's stated positions and bad ideas \cite{cheng2026, sharma2024sycophancy}, because there is no objective way to score whether a stranger's career choice is a bad idea. Within what we could measure, a behavior the laboratory literature warns about ran at full strength inside a wellbeing intervention and did no detectable harm.

That leaves the explanation our data do support: decision pressure. The script told the agent to get a decision at each step, and the agent obeyed. When participants hesitated, it asked again. Nearly half of its requests to decide came bundled with praise for one of the options, one request in seven was repeated after the participant had declined to choose, and the commitments produced by those repeats were the weakest in the corpus. The participants who fielded the most requests ended the study most doubtful about their plans, which is what self-determination theory predicts: a choice a person is talked into is held less firmly than one they arrive at on their own \cite{ryan2000}. Work on decision mindsets and closure pressure \cite{gollwitzer1990, webster1994} adds a related point: when people are pushed to decide before they have finished weighing their options, they still decide, but on less thinking than the decision needed. One participant summed the experience up: ``it would be useful probably to reflect on other job options too. For this study I had to choose one direction.'' The journaling arm answered the same decision questions without harm, so the difference between the arms comes down to who is asking. A form accepts a hedged answer and moves on. An agent waits, follows up and asks again, and people respond to a machine that behaves socially as if it were someone rather than something \cite{reeves1996, nass2000}.

\textbf{Give the agent a decision-demand budget.} Asking participants to decide was the only agent behavior linked to a worse outcome. The total number of demands carried the association, and the re-asks are where demands visibly produced weaker commitments (Section~\ref{sec:rq3}). We suggest three rules, enforceable by checking each reply before it is sent, without changing the model. First, ask for a decision once: if the participant responds with a mix of options, with ``I'm not sure'' or with a wish to decide later, that counts as an answer, because asking a second time produced weaker commitments than waiting. Second, do not praise an option in the same turn that asks the participant to choose, because praise attached to a question turns it into a nudge. Third, save the wrap-up for the end: pulling choices together into a plan can happen in a written summary at the close of the session, which is how the journaling version did it without harm. Anyone with the transcript can see whether each rule was followed.

\textbf{Put the insight material where you want the insight.} Participants said insightful things about themselves almost only on the first day, when the script had them talking about their own past; on the planning days, insight nearly disappeared (Section~\ref{sec:rq2}). If a program wants insight throughout, the material that draws people back to their own story needs to appear on every day rather than only the first. Our data cannot prove that spreading it out works, only that front-loading it left three days nearly empty of insight.

One thing the results cannot fully explain is the rest of the trial's pattern. Our decision rule in Section~\ref{sec:analytic} anticipated a clean verdict, and the data returned a mixed one: cumulative decision demands predicted post-test doubt, and nothing else at either level predicted any outcome. For everything except doubt, whatever separated the agent condition from journaling must be something every agent conversation shared, such as answering to a social counterpart rather than to a page. Our codes count how often behaviors happen, and a constant present in every conversation is invisible to counts. Identifying that shared ingredient is the most important question this work leaves open.

Our findings also help the trial: if the agent had behaved erratically, that could have explained the trial's result, but it did not: the agent's behavior varied a lot from participant to participant, and apart from decision demands, none of that variation predicted how anyone fared. The trial also proposed a reason for its result that it could not test: drawing on self-determination theory, it suggested that writing alone leaves people feeling free and in control of their reflections in a way that answering to an agent does not \cite{soh2026tmb}. Our transcripts show what that looks like in practice. The agent repeatedly asked participants to decide, and when they hesitated it asked again, taking away the option of not deciding. Finally, the trial suggested that future prompts should ask the agent to offer alternative perspectives and probing questions. Our fidelity result speaks directly to that suggestion: the prompt already asked for gently challenging questions and the agent almost never delivered them (Section~\ref{sec:rq1}). Changing the wording alone would probably not have been enough.

\subsection{Limitations}
This work has several limitations. Career doubt, the outcome our central association concerns, was an exploratory outcome in the trial rather than a pre-registered one, so our transcript measures connect to an exploratory trial finding. The daily surveys measured general identity states rather than career-specific ones, which may be part of why no day-level association appeared. Our central association, between decision demands and doubt, is a single correlation drawn from 32 tests on 107 people. It survived every check we ran, but one version of the analysis leaves it close to the significance line, and the demands were not randomly assigned, so a participant's own indecision could still explain part of it. That is why we call it a candidate rather than a cause. The null results have limits of their own: the day-level tests could have missed effects smaller than about .10, and because every participant received at least some praise, we cannot say what receiving none would do. Challenge was the code our human coders agreed on least, so the challenge findings lean on the annotator we validated for it. The annotators were validated on Study 2, and the Study 1 labels are used descriptively. Both studies used one model (GPT-4o), one topic (careers) and United States samples, and the trial's effects were small to begin with, so the mechanisms here explain modest differences. Finally, these findings describe GPT-4o as deployed in 2025; Section~\ref{sec:promptimplications} discusses which parts of the picture we expect to survive model change and why the audit itself is how anyone would find out.

\section{Conclusion}

A randomized trial found that emerging adults who reflected on their careers with a conversational agent ended the program less committed to their plans than those who worked through the same questions in a journaling interface, and we examined all 17,930 conversation turns to understand why. Under two different prompts, every rule the agent followed was one that could be checked, and no rule about how to behave was followed. Its conversations closed well, and its day-to-day behavior made no detectable difference to how participants felt. The behavior linked to the added doubt was the demand to decide: the journaling format posed each decision once on the page, while the agent posed it and, when a participant hesitated, posed it again. Designers of reflection agents should budget decision demands, let participants decline to decide, specify behavior in verifiable terms and audit transcripts as a matter of routine. Whether decision pressure causes doubt is now a precise and testable question and we hope someone runs the experiment.

\section*{Disclosure of AI Use}
We used large language models in this research as annotation instruments (described and validated in Section 4) and as assistance in preparing and revising the analysis code and the manuscript. We designed, verified and approved all analyses, analytic decisions and claims, and we reviewed and edited all manuscript text.


\bibliographystyle{ACM-Reference-Format}
\bibliography{references}

\appendix

\section{The Two System Prompts}
\label{app:prompts}

Each prompt repeats the same core instructions for all four activities, changing only the activity's topic list. We reproduce the core instructions of each here, lightly abridged; the complete deployed text of both prompts, including the topic lists and safety scripts for every activity, is in supplementary materials.

\subsection*{Study 1 (``facilitator'')}

\textit{Role.} ``You are a career reflection facilitator for emerging adults in college. Your role is to guide thoughtful, emotionally supportive conversations that promote career reflection.''

\textit{Golden rules.} ``Only ask one question at a time to not overwhelm the user. Make sure user shares AT LEAST 15 responses. Always perform RESPONSE MONITORING immediately after receiving a user response. Ensure all ACTIVITY GOALS are met.''

\textit{Core rules.} ``Ask relevant, personalized follow-up questions based on what the user shares. Keep the conversation on-topic to career goals, coherent, and evolving, building on previous reflections. Offer new suggestions or insights based on what the user shares. Distill, connect, and synthesize key insights from across conversations. Be emotionally attentive and supportive of the user's experiences and feelings, without being overly positive.''

\textit{Conversation style.} ``Keep your tone warm, curious, and emotionally present---like a thoughtful peer who is actively listening. Do not overdo positive feedback---only validate when it feels necessary or appropriate (e.g., not after every user response). Avoid providing long responses---keep it concise, natural, and user-focused. To conclude conversation after ACTIVITY GOALS are met, thank the participant for sharing their reflections. Summarize and synthesize the key, topically relevant insights.''

\textit{Response monitoring.} Each user response is classified (short, lacking, topically inconsistent, off-topic, noncompliant, or crisis) and the prompt scripts a specific reaction to each class, including fixed wording for crisis disclosures.

\subsection*{Study 2 (``conversation partner'')}

\textit{Role.} ``Your role is to be a conversation partner who helps the user reflect on their career path by asking thoughtful questions and engaging in a natural, two-way conversation.''

\textit{Main goals.} ``Cover all topics in this activity, focusing only on the provided questions in order. Be natural, flexible, and conversational---avoid seeming scripted. Use a tone that is friendly and approachable, ranging from neutral to warm, while conveying understanding and support. Do not be overly agreeable or automatically validate everything the user says. Keep your replies short (1--2 sentences). Encourage deep reflection on the user's reasoning, feelings, and experiences related to their career path. Ask open-ended, clarifying, and gently challenging follow-up questions. Keep the conversation moving and avoid lingering too long on a single topic. Occasionally summarize user's insights when helpful, highlighting patterns, themes, or strengths. Once all activity topics are covered, always conclude with the ENDING.''

\textit{Response monitoring.} Fixed scripts cover crisis disclosures, requests to stop early and off-topic turns; in all other cases the agent is told to ask ``open-ended, clarifying, and gently challenging questions'' with one to two follow-ups per topic.

\textit{Ending.} Before closing, the agent must ``ask the user to summarize the main themes from this conversation''; it then thanks the user and posts the activity's completion code.

\section{The Annotator Prompt}
\label{app:annotatorprompt}

Every annotator model received the same fixed prompt at temperature 0. The prompt has three parts: a role instruction, the full codebook and a set of worked examples. The role instruction for agent turns reads:

\begin{quote}\small
``You are a trained content-analysis coder annotating ASSISTANT turns from an AI career-reflection intervention, following the codebook below exactly. Mark every move present (multi-label). Respond with ONLY a JSON object with integer 0/1 values for keys: q\_open, q\_closed, recap\_synth, validation, challenge, advice, process, closure\_move\_01, stacked\_q\_01.''
\end{quote}

The participant-turn instruction is analogous: it asks for one primary category, a certainty rating from 1 to 5 (or NA when the turn takes no career-relevant stance) and a 0/1 insight flag, returned as a JSON object. After the instruction, the prompt includes the same codebook document the human coders used, followed by the worked examples: turns on which both human coders fully agreed, each shown with its conversational context and its agreed labels. Each turn to be coded is then presented in the same format as the examples, as CONTEXT (the two preceding turns), TURN (the text to code) and LABELS, which the model completes. The full codebook, with definitions, decision rules and all worked examples, is in supplementary materials.

\section{Annotator Selection}
\label{app:annotators}

Table~\ref{tab:models} reports the validation of all four candidate annotators against the consensus gold standard on held-out items. We accessed the models through an API broker at temperature 0 with a frozen prompt: anthropic/claude-sonnet-5, google/gemini-flash (latest, August 2026), openai/gpt-5.6-luna-pro and x-ai/grok-4.5. The selection rule, fixed in advance, was to deploy an annotator matching or exceeding human reliability for each field, preferring the less expensive model when several qualified.

\begin{table}[H]
\caption{Candidate annotator agreement with the consensus gold standard ($\kappa$, with certainty as MAE). Bold marks the annotator of record for each field.}
\label{tab:models}
\scriptsize\setlength{\tabcolsep}{2pt}
\begin{tabular}{lcccc}
\toprule
Field & Sonnet~5 & Flash & Luna & Grok\\
\midrule
Open question & .70 & 1.00 & \textbf{.87} & .97\\
Closed question & .77 & .94 & \textbf{.94} & .97\\
Restate / synth. & .93 & .97 & \textbf{.97} & .85\\
Validation & .86 & .86 & \textbf{.95} & .97\\
Challenge & \textbf{.85} & .32 & .20 & .39\\
Advice & .85 & .79 & \textbf{.85} & 1.00\\
Process & .92 & 1.00 & \textbf{.97} & 1.00\\
Closure move & .86 & .96 & \textbf{.85} & .97\\
Question stacking & .71 & 1.00 & \textbf{.94} & .85\\
\hdashline
Participant category & .70 & .83 & \textbf{.81} & .84\\
Insight & \textbf{.50} & .17 & .42 & .33\\
Certainty (MAE) & .56 & .35 & \textbf{.32} & .30\\
\bottomrule
\end{tabular}
\end{table}

\section{Day-Level Models}
\label{app:daylevel}

Table~\ref{tab:daylevel} reports the standardized within-person coefficient for every predictor-outcome pair, estimated from 448 session-days with linear mixed models that include the predictor's person mean, activity fixed effects, the outcome's baseline-week mean where available and random intercepts. No coefficient is significant at $p<.05$ before correction and none survives it.

\begin{table}[H]
\caption{Within-person standardized coefficients, each day's conversation features predicting that evening's state.}
\label{tab:daylevel}
\footnotesize\setlength{\tabcolsep}{2.6pt}
\begin{tabular}{lcccccc}
\toprule
 & Commit & Rumin. & Aff+ & Aff-- & Insight & Career ins.\\
\midrule
Certainty & $-$.03 & $-$.01 & +.02 & +.03 & +.07 & $-$.02\\
Validation & $-$.00 & $-$.03 & +.01 & +.02 & +.02 & +.03\\
Stacking & +.03 & +.00 & +.03 & $-$.07 & +.07 & $-$.01\\
Agent words & $-$.07 & $-$.02 & +.05 & +.05 & +.10 & +.07\\
Closure moves & +.04 & +.02 & $-$.04 & $-$.01 & $-$.13 & +.03\\
Repeated demands & +.04 & +.01 & +.01 & $-$.01 & $-$.06 & $-$.01\\
Any challenge & +.03 & $-$.01 & $-$.03 & +.03 & $-$.03 & +.02\\
Commitment talk & $-$.01 & $-$.06 & +.03 & $-$.01 & +.04 & +.08\\
\bottomrule
\end{tabular}
\end{table}

\section{Person-Level Models}
\label{app:personlevel}

Table~\ref{tab:personlevel} reports standardized coefficients from regressions of each post-test outcome on each conversation feature (n=107; controls for the outcome's baseline scale, age and gender; robust standard errors). Only closure moves predicting career doubt survives Benjamini-Hochberg correction across the 32 tests. As a sensitivity check on that association, an unmeasured confounder would need associations of roughly 2.6 in risk-ratio terms (an E-value \cite{vanderweele2017}) with both closure exposure and doubt, beyond baseline doubt and demographics, to fully explain the estimate.

\begin{table}[H]
\caption{Person-level standardized coefficients, cumulative conversation features predicting post-test outcomes. $^{*}p<.05$, $^{**}p<.01$ uncorrected; $^{\dagger}$ survives FDR correction.}
\label{tab:personlevel}
\footnotesize\setlength{\tabcolsep}{2.6pt}
\begin{tabular}{lcccc}
\toprule
 & Commit & Career com. & Doubt & Rumin.\\
\midrule
Validation rate & +.01 & $-$.00 & +.04 & +.05\\
Challenge rate & +.00 & +.06 & $-$.08 & $-$.18$^{*}$\\
Stacking rate & +.03 & +.09 & $-$.00 & $-$.03\\
Closure moves & +.13$^{*}$ & +.03 & +.25$^{**\dagger}$ & +.14$^{**}$\\
Repeated demands & +.09 & $-$.07 & +.12 & +.03\\
Certainty & $-$.02 & +.10 & $-$.03 & +.03\\
Commitment talk & +.05 & +.14 & $-$.16$^{*}$ & $-$.15$^{*}$\\
Insight count & +.08 & +.14 & $-$.07 & $-$.06\\
\bottomrule
\end{tabular}
\end{table}

\end{document}